\documentstyle[12pt,twoside,titlepage,epsf]{article}  
\newcommand{\ncm}{\newcommand}
\renewcommand{\theequation}{\thesection.\arabic{equation}}
\newcommand{\sectiona}[1]{\setcounter{equation}{0}\section{#1}}
\ncm{\oH}{\bar{H}}
\ncm{\us}{\quad\mbox{using}\quad}
\ncm{\ra}{\rightarrow}
\ncm{\ot}{\otimes}
\ncm{\DH}{D(H)}
\ncm{\DW}{D^{\omega}(H)}
\ncm{\TH}{T(\oH)}
\ncm{\ssc}{\displaystyle}
\ncm{\oq}{\eta} 
\ncm{\im}{\imath}
\ncm{\ba}{\begin{array}}
\ncm{\ea}{\end{array}}
\ncm{\ul}{\underline}
\ncm{\ol}{\overline}

\ncm{\str}{\rule{0cm}{3.5mm}}
\ncm{\om}{\omega}
\ncm{\ep}{\epsilon}
\newlength{\extraspace}

\newlength{\extraspaces}
\newcommand{\be}{\begin{equation}
\addtolength{\abovedisplayskip}{\extraspaces}
\addtolength{\belowdisplayskip}{\extraspaces}
\addtolength{\abovedisplayshortskip}{\extraspace}
\addtolength{\belowdisplayshortskip}{\extraspace}}
\newcommand{\ee}{\end{equation}}

\makeatletter

\def\eqnfourarray{\stepcounter{equation}%
  \def\@currentlabel{\p@equation\theequation}\global\@eqnswtrue \m@th
  \global\@eqcnt\z@ \tabskip\@centering
  \let\\\@eqncr \let\@@eqncr\@@eqnfourcr
  $$\everycr{}\halign to\displaywidth \bgroup
  \hskip\@centering $\displaystyle \tabskip\z@skip {##}$\@eqnsel &%
  \global\@eqcnt\@ne \hskip\tw@\arraycolsep \hfil ${##}$\hfil &%
  \global\@eqcnt\tw@ \hskip\tw@\arraycolsep $\displaystyle {##}$\hfil &%
  \global\@eqcnt\thr@@ \hskip\tw@\arraycolsep \hfil \hbox{##}%
    \tabskip\@centering &%
  \global\@eqcnt 4 \hbox\bgroup \hss ##\egroup
    \tabskip\z@skip \cr}

\def\endeqnfourarray{\@@eqncr \egroup
  \global\advance\c@equation\m@ne $$\global\@ignoretrue}

\def\@@eqnfourcr{\let\reserved@a\relax
  \ifcase\@eqcnt \def\reserved@a{& & & &}%
    \or \def\reserved@a{& & &}%
    \or \def\reserved@a{& &}%
    \or \def\reserved@a{&}%
    \else \let\reserved@a\@empty
      \@latex@error{Too many columns in eqnfourarray environment}\@ehc
  \fi
  \reserved@a \if@eqnsw\@eqnnum \stepcounter{equation}\fi
  \global\@eqnswtrue \global\@eqcnt \z@\cr}

\makeatother

\newcommand{\bea}{\begin{eqnarray}
\addtolength{\abovedisplayskip}{\extraspaces}
\addtolength{\belowdisplayskip}{\extraspaces}
\addtolength{\abovedisplayshortskip}{\extraspace}
\addtolength{\belowdisplayshortskip}{\extraspace}}
\newcommand{\eea}{\end{eqnarray}}
\newcommand{\beas}{\begin{eqnarray*}
\addtolength{\abovedisplayskip}{\extraspaces}
\addtolength{\belowdisplayskip}{\extraspaces}
\addtolength{\abovedisplayshortskip}{\extraspace}
\addtolength{\belowdisplayshortskip}{\extraspace}}
\newcommand{\eeas}{\end{eqnarray*}}
\ncm{\Z}{{\mbox{\bf Z}}}
\ncm{\al}{\alpha}
\ncm{\bt}{\beta}
\ncm{\gm}{\gamma}
\ncm{\dl}{\delta}
\ncm{\varep}{\varepsilon}
\ncm{\zt}{\zeta}
\ncm{\et}{\eta}
\ncm{\th}{\theta}
\ncm{\kp}{\kappa}
\ncm{\lm}{\lambda}
\ncm{\rh}{\rho}
\ncm{\hl}{\hline}
\ncm{\sg}{\sigma}
\ncm{\ta}{\tau}
\ncm{\ph}{\phi}
\ncm{\phv}{\varphi}
\ncm{\ch}{\chi}
\ncm{\ps}{\Phi}
\ncm{\nn}{\nonumber}
\title{ 
{\flushleft {\normalsize April 1997
\hfill PAR--LPTHE 97--11, hep-th/9704063}}\\[2.5cm]
CONFINEMENT IN PARTIALLY BROKEN  \\ 
ABELIAN CHERN-SIMONS THEORIES \vspace{1.5cm} }
\author{ 
Mark de Wild Propitius\thanks{e-mail: mdwp@lpthe.jussieu.fr}\\
\normalsize{ {\em
Laboratoire de Physique Th\'eorique et Haute Energies}}\thanks{Laboratoire 
associ\'e No.\ 280 au CNRS}\\
\normalsize{{\em Universit\'e Pierre et Marie Curie - PARIS VI}} \\
\normalsize{{\em Universit\'e Denis Diderot - PARIS VII}} \\
\normalsize{{\em 4 place Jussieu, Boite 126, Tour 16, 1$^{er}$ \'etage}}\\
\normalsize{{\em F-75252 Paris CEDEX 05, France}}
\vspace{6cm}} 
\date{}
\begin{document}
\maketitle
\begin{abstract}
Planar Chern-Simons (CS) theories in which a compact abelian gauge group 
$U(1) \times U(1)$ is spontaneously broken to $U(1) \times \Z_N$ 
are investigated. Among other things, it is noted that the theories just 
featuring the mixed CS term coupling the broken to the unbroken 
$U(1)$ gauge fields in general exhibits an interesting form of confinement: 
only particles carrying certain multiples of the fundamental magnetic 
vortex flux 
unit and certain multiples of the fundamental charge of the unbroken 
$U(1)$ gauge field can appear as free particles. Adding the 
usual CS term for the broken $U(1)$ gauge fields does not change much. 
It merely leads to additional Aharonov-Bohm interactions among these 
particles. Upon introducing the CS term 
for the unbroken $U(1)$ gauge fields, in contrast, the  
confinement phenomenon completely disappears.

\end{abstract}
\newpage

\sectiona{Introduction}

The spectrum and the Aharonov-Bohm 
interactions described by all conceivable 2+1 dimensional 
Chern-Simons theories in which every compact $U(1)$ factor of a direct 
product gauge group $U(1)^k$ are broken down to a finite 
cyclic subgroup by means of the Higgs mechanism have recently been 
established in~\cite{multiCS}. Although briefly mentioned,
the interesting possibility that some $U(1)$ gauge groups
remain unbroken was not explored there. The purpose of this paper 
is to give a complete description of these partially broken theories.
We will exclusively work in 
2+1 dimensional Minkowski 
space with signature $(+,-,-)$. Greek indices run from 0 to 2 and spatial 
components are labeled by Latin indices.
Natural units in which $\hbar=c=1$ are used throughout.

\sectiona{The model}

For convenience, we will focus on the simplest example of a 
partially broken 2+1 dimensional 
abelian Chern-Simons (CS) theory, namely  that in which 
the compact gauge group 
$U(1) \times U(1)$ is spontaneously broken down to $U(1) \times \Z_N$.     
The most general action we can write down for such a theory 
is of the form
\bea        \label{action}
S &=&  \int d\,^3x \;  ({\cal L}_{\mbox{\scriptsize YMH}} + 
{\cal L}_{\mbox{\scriptsize matter}}
+ {\cal L}_{\mbox{\scriptsize CS}}) \, ,  \\
{\cal L}_{\mbox{\scriptsize YMH}} &=& 
-\frac{1}{4}F^{(i)\kappa\nu} F^{(i)}_{\kappa\nu} +
({\cal D}^\kappa \Phi)^*{\cal D}_\kappa \Phi - V(\Phi) \, ,  \\
{\cal L}_{\mbox{\scriptsize matter}} &=& 
- j^{(i)\kappa}A^{(i)}_{\kappa} \, ,
\label{mat} \\
{\cal L}_{\mbox{\scriptsize CS}} &=& 
\frac{\mu^{(i)}}{2} \epsilon^{\kappa\nu\tau} 
A^{(i)}_{\kappa} \partial_{\nu}A^{(i)}_{\tau} +
\frac{\mu^{(12)}}{2} \epsilon^{\kappa\nu\tau} 
A^{(1)}_{\kappa} \partial_{\nu}A^{(2)}_{\tau} \, ,    \label{CS}
\eea 
where $A^{(i)}_\kappa$ with $i=1,2$ denote two compact $U(1)$ gauge fields 
with coupling constant  $e^{(i)}$. 
All repeated Greek and Latin indices are summed over. 
Except where otherwise stated  this summation convention 
holds for the rest of the paper. The Higgs field  $\Phi$ carries 
charge $N e^{(2)}$, i.e.\ ${\cal D}_\kappa \Phi = 
(\partial_{\kappa}+\im N e^{(2)} A_{\kappa}^{(2)})\Phi$. It is endowed 
with a nonvanishing vacuum expectation value $v$ through the well-known 
potential $V(\Phi) = \frac{\lambda}{4}(|\Phi|^2-v^2)^2$  
with $\lambda, v > 0$. Thus the compact gauge group $U(1) \times U(1)$ 
is spontaneously broken down to $U(1) \times \Z_N$ at the energy scale
$M_H = v \sqrt{2\lambda}$. To proceed, the charges introduced by the 
matter currents $j^{(1)}$ and $j^{(2)}$ in~(\ref{mat}) are quantized
as $q^{(1)} = \int\! d\, ^2x \, j^{(1) \; 0} = n^{(1)} e^{(1)}$ 
and $q^{(2)}= \int\! d\, ^2x \, j^{(2) \; 0} = n^{(2)} e^{(2)}$ 
with $n^{(1)},n^{(2)} \in \Z$. We will also assume the presence 
of Dirac monopoles which implies that the topological masses in~(\ref{CS})
satisfy the quantization conditions (see~\cite{multiCS} and references 
therein) 
\bea                                  
\mu^{(i)} \;=\; \frac{p^{(i)} e^{(i)}e^{(i)}}{\pi} 
\;\;\; \mbox{and} \;\;\; 
\mu^{(12)} \;=\; \frac{p^{(12)} e^{(1)}e^{(2)}}{\pi} 
\;\;\; \mbox{with $p^{(i)},
p^{(12)} \in {\mbox{\bf Z}}$} \,.       
\label{quantmu}  
\eea   
As an exception to the rule, there is no summation over the repeated 
index  $(i)$ in this case.

\subsection{$\mu^{(1)} \neq 0$ }

In the following, the main emphasis will be on the effective low energy 
($E \ll M_H$) or equivalently the effective long distance ($r \gg 1/M_H$)  
physics described by~(\ref{action}). We will first take 
the topological mass $\mu^{(1)}$ to be nonzero and return to the special case 
$\mu^{(1)}=0$ in section~\ref{mu0}. 

In the low energy regime the Higgs field $\Phi$ takes groundstate values 
everywhere, i.e.\  $\Phi(x)= v \exp(\im \sigma(x))$. Thus the  
low energy action is obtained by the following replacement in~(\ref{action}) 
\bea                               \label{efhi}
{\cal L}_{\mbox{\scriptsize YMH}} & 
\longmapsto & -\frac{1}{4}F^{(i) \kappa\nu} F^{(i)}_{\kappa\nu} 
+\frac{M_A^2}{2} \tilde A^{\kappa}\tilde A_{\kappa} \, , 
\eea
with $\tilde{A}_{\kappa} := A_{\kappa}^{(2)} + 
\partial_{\kappa} \sigma / {Ne^{(2)}}$ and 
$M_A  :=  Ne^{(2)} v\sqrt{2}$.
Varying this effective low energy action w.r.t.\ the two gauge fields 
yields the field equations 
\bea 
\partial_\nu F^{(1)\nu\kappa} 
&=& j^{(1) \kappa} 
- \mu^{(1)}\epsilon^{\kappa\nu\tau} \partial_{\nu}A^{(1)}_{\tau} 
-\frac{\mu^{(12)}}{2}\epsilon^{\kappa\nu\tau} \partial_{\nu}A^{(2)}_{\tau} 
\, , \label{feq1} \\
\partial_\nu F^{(2) \nu\kappa} 
&=& 
j^{(2)\kappa}
- \mu^{(2)}\epsilon^{\kappa\nu\tau} \partial_{\nu}A^{(2)}_{\tau} 
- \frac{\mu^{(12)}}{2}\epsilon^{\kappa\nu\tau} \partial_{\nu}A^{(1)}_{\tau} 
+j^\kappa_{\mbox{\scriptsize scr}}  
\, ,                       
\label{feq2}
\eea
with $j^\kappa_{\mbox{\scriptsize scr}} := - M_A^2 \tilde{A}^\kappa$.
It is easily verified that these equations imply that both gauge fields 
in this CS Higgs medium are massive.

There are three independent particle-like sources in this theory, 
namely the quantized matter charges $q^{(1)}=  n^{(1)} e^{(1)}$ and 
$q^{(2)}=  n^{(2)} e^{(2)}$ and the magnetic vortices corresponding
to topologically stable finite energy solutions. In the low energy 
regime such a field configuration can be idealized as a 
point ${\bf x}_0$ in the plane where the Higss field vanishes: 
$\Phi({\bf x}_0)=0$. Around this point the Higgs field makes a 
noncontractible winding in the vacuum manifold. That is, 
$\Phi({\bf x})= v \exp(\im \sigma({\bf x}))$  
for ${\bf x} \neq {\bf x}_0$ and  
$\oint_\gamma  dl^i\partial_i \sigma=2\pi a$ with $\gamma$ a  
loop enclosing ${\bf x}_0$ and $a \in \Z$ to the render Higgs field
single valued. For finite energy, the covariant derivative of the Higgs 
field should vanish away from ${\bf x}_0$, i.e.\
\bea \label{phova}
{\cal D}_i \Phi({\bf x} \neq {\bf x}_0)=0 &\Longrightarrow& 
\tilde{A}_i({\bf x} \neq {\bf x}_0)=0 \, . 
\eea 
So the holonomy in the Goldstone boson field $\sigma$ is 
accompanied by a holonomy in the gauge field $A^{(2)}_\kappa$. The 
well-known  conclusion is that the 
vortices carry a quantized magnetic flux $\phi^{(2)}  
=  \oint_\gamma dl^i A^{(2)\,i} =   
\oint_\gamma  dl^i\partial_i \sigma / {Ne^{(2)}}
 =  {2\pi a}/{Ne^{(2)}}$, 
with $a \in \Z$, located at 
some point ${\bf x}_0$  in the plane.

Both the matter charges and the magnetic vortices enter the effective field 
equations~(\ref{feq1}) and~(\ref{feq2})  describing the physics  
away from the locations of the vortices. 
The matter charges enter by means of the 
matter currents $j^{(1)\,\kappa}$ and $j^{(2)\,\kappa}$ and the vortices 
through the magnetic flux current 
$-\frac{1}{2}\epsilon^{\kappa\nu\tau} \partial_{\nu}A^{(2)}_{\tau}$.
From these equations we learn that 
both these matter currents and this flux current generate electromagnetic 
fields, which are screened at large distances by an 
induced appropriate combination of the screening current 
$j^\kappa_{\mbox{\scriptsize scr}}$ 
and the flux current  
$-\frac{1}{2}\epsilon^{\kappa\nu\tau} \partial_{\nu}A^{(1)}_{\tau}$.
To be specific, the Gauss' laws following from~(\ref{feq1}) 
and~(\ref{feq2}) read 
\bea     
Q^{(1)} &=& q^{(1)} + \mu^{(1)} \phi^{(1)} + 
\frac{\mu^{(12)}}{2} \phi^{(2)} \; = \; 0  \, ,  \label{g1}   \\
Q^{(2)} &=& q^{(2)}  + \mu^{(2)} \phi^{(2)} +
\frac{\mu^{(12)}}{2} \phi^{(1)} +q_{\mbox{\scriptsize scr}} 
\; = \; 0 \label{g2} \, ,
\eea  
with $Q^{(i)}: =  \int\! d\, ^2x\, \partial_j F^{(i)\, j0}$
the Coulomb charges which both vanish since both $U(1)$ gauge 
fields are massive. The Gauss' law~(\ref{g1}) indicates that the long range 
Coulomb fields  $F^{(1)\, j0}$ generated by a particle
$(q^{(1)}, q^{(2)},  \phi^{(2)})$ being a composite of the quantized
matter charges $q^{(1)}$ and $q^{(2)}$ and a vortex  $\phi^{(2)}$ 
are screened by an induced screening flux $\phi^{(1)}$. 
From~(\ref{g2}), we subsequently infer that the long range 
Coulomb fields  $F^{(2)\, j0}$ generated by this screening flux $\phi^{(1)}$ 
and the charge $q^{(2)}$ and flux $\phi^{(2)}$ carried by this particle
are screened by a screening charge $q_{\mbox{\scriptsize scr}} := 
\int \! d\,^2 x \, j_{\mbox{\scriptsize scr}}^0$ induced 
in the Higgs condensate~\cite{screening}. 
Note that although the matter charges $q^{(1)}$ and $q^{(2)}$ and 
the magnetic flux $\phi^{(2)}$ are necessarily quantized in this
spontaneously broken compact gauge theory, both the screening flux
$\phi^{(1)}$ and screening charge $q_{\mbox{\scriptsize scr}}$ can 
in principle take any real value. In other words, for each particle 
$(q^{(1)}, q^{(2)},  \phi^{(2)})$ in the spectrum of this theory 
there always exists a screening flux
$\phi^{(1)}$ and screening charge $q_{\mbox{\scriptsize scr}}$
such that the long range Coulomb fields indeed vanish. 
As we will see in section~\ref{mu0}, for $\mu^{(1)}=0$ this is no longer 
the case.

Since the long range electromagnetic fields are completely screened 
there are no classical long range interactions among the particles   
$(q^{(1)}, q^{(2)},  \phi^{(2)})$. The long range interactions we are 
left with are the quantum mechanical Aharonov-Bohm (AB) interactions due
to the matter couplings~(\ref{mat}) and the CS 
couplings~(\ref{CS}). Specifically, a 
counterclockwise monodromy ${\cal R}^2$ of a particle 
$(q^{(1)}, q^{(2)},  \phi^{(2)})$ and a remote particle  
$(q^{(1)'}, q^{(2)'},  \phi^{(2)'})$ leads to the  
AB phase (e.g.\ \cite{multiCS})
\bea \label{mono}
{\cal R}^2 &=& \exp \left( \im q^{(i)}\phi^{(i)'} + \im q^{(i)'}\phi^{(i)}  
+ \im \mu^{(i)} \phi^{(i)}\phi^{(i)'}  
+\im \frac{\mu^{(12)}}{2} \left( \phi^{(1)}\phi^{(2)'}+ 
\phi^{(1)'}\phi^{(2)} \right)
\right), \qquad
\eea 
whereas a counterclockwise braiding  of two remote identical 
particles $(q^{(1)}, q^{(2)},  \phi^{(2)})$ gives rise to 
the AB phase  
\bea \label{braid}
{\cal R} &=&
\exp \left( \im q^{(i)}\phi^{(i)} + \im \mu^{(i)} \phi^{(i)}\phi^{(i)} + 
\im \frac{\mu^{(12)}}{2}\phi^{(1)}\phi^{(2)} \right).
\eea 
A couple of remarks are pertinent here. First of all, for each particle 
$(q^{(1)}, q^{(2)},  \phi^{(2)})$ in the spectrum there  
exists an anti-particle $(\bar{q}^{(1)}, \bar{q}^{(2)}, \bar{\phi}^{(2)})$
such that the pair may decay into the vacuum. In other words, the topological 
proof of the spin-statistics connection (see for instance~\cite{multiCS} 
and references given there) applies to this theory. Thus the quantum 
statistics phase~(\ref{braid}) is the same as the spin factor obtained 
by a counterclockwise rotation over an angle of $2\pi$ of the particle
$(q^{(1)}, q^{(2)},  \phi^{(2)})$. Secondly, a crucial 
observation~\cite{screening} 
in the derivation of the AB phases~(\ref{mono}) 
and~(\ref{braid}) is that in contrast to the screening fluxes 
$\phi^{(1)}$ the screening charges $q_{\mbox{\scriptsize scr}}$ 
attached to the particles do not couple to the AB 
interactions.  The point is that the screening charges 
$q_{\mbox{\scriptsize scr}}$ do not only  couple to the holonomy in the 
gauge field $A^{(2)}_\kappa$ around a remote vortex 
$\phi^{(2)}$, but also to the holonomy 
in the Goldstone boson field $\sigma$. This is immediate from~(\ref{efhi}).
Let $j_{\mbox{\scriptsize scr}}^\kappa := - M_A^2 \tilde{A}^\kappa$ be 
the screening current associated with some screening charge
$q_{\mbox{\scriptsize scr}}$. The second term at the l.h.s.\ 
of~(\ref{efhi}) couples this current to the combined field
$\tilde{A}_\kappa$ around the remote vortex:  
$j_{\mbox{\scriptsize scr}}^\kappa \tilde{A}_\kappa$.
As we have seen in~(\ref{phova}), away from the location ${\bf x}_0$ of 
the vortex the holonomies in the gauge fields and the Goldstone boson are
related such that $\tilde{A}_\kappa$ strictly vanishes. 
Consequently, as long as the screening charge stays away from the location
of the vortex, the interaction term 
$j_{\mbox{\scriptsize scr}}^\kappa \tilde{A}_\kappa$
vanishes and therefore does not generate an AB phase in the process
of taking a screening charge around a remote vortex.     

Henceforth, the 
particles  $(q^{(1)} = n^{(1)} e^{(1)}, q^{(2)} = n^{(2)} e^{(2)},
\phi^{(2)} = {2\pi a}/{Ne^{(2)}})$  will be labeled 
as $(n^{(1)}, n^{(2)}, a)$.
In terms of these integral quantum numbers the AB 
phases~(\ref{mono}) and~(\ref{braid}) become
\bea 
{\cal R}^2  &=& \exp \left(
 \frac{2\pi \im}{N} ( n^{(2)}a' + n^{(2)'}a + \frac{2p^{(2)}}{N}aa') 
 -\frac{\pi \im}{p^{(1)}}(n^{(1)}+\frac{p^{(12)}}{N} a)
 (n^{(1)'}+\frac{p^{(12)}}{N} a')   
\right), \qquad  \;     \label{monq}  \\
{\cal R}  &=& \exp \left(
 \frac{2\pi \im}{N} ( n^{(2)}a + \frac{p^{(2)}}{N}aa) 
 -\frac{\pi \im}{2p^{(1)}}(n^{(1)}+\frac{p^{(12)}}{N} a)
 (n^{(1)}+\frac{p^{(12)}}{N} a)   
\right),  \label{braidq}
\eea 
where we substituted the values of the 
screening fluxes $\phi^{(1)}$ and $\phi^{(1)'}$ 
following from  the Gauss' law~(\ref{g1}) and  the quantization 
of the topological masses~(\ref{quantmu}).
Note that under these long range AB 
interactions the integral charge label $n^{(2)}$ becomes 
a $\Z_N$ quantum number.

Let us now turn to the Dirac monopoles that may be introduced 
in this compact gauge theory. There are two species carrying the 
quantized magnetic charges $g^{(1)}= {2\pi m^{(1)}}/{e^{(1)}}$ 
and $g^{(2)}= {2\pi m^{(2)}}/{e^{(2)}}$ with 
$m^{(1)}, m^{(2)} \in \Z$. In this 2+1 dimensional Minkowski setting 
these monopoles are instantons tunneling between states with flux 
difference $\Delta \phi^{(1)} = {2\pi m^{(1)}}/{e^{(1)}}$ and 
$\Delta \phi^{(2)} = -{2\pi m^{(2)}}/{e^{(2)}}$. From the 
Gauss' laws~(\ref{g1}) and~(\ref{g1})  we infer that 
these flux tunnelings are accompanied by charge tunnelings. 
Specifically, in terms of our favourite   
integral charge and flux labels the tunnelings induced by the 
two distinct minimal monopoles read~\footnote{It should be noted that   
the Dirac monopoles in this CS theory actually only appear as 
monopole/anti-monopole pairs linearly confined by a string representing 
the wordline of the particle created/annihilated by the 
monopole/anti-monopole in these pairs~\cite{pisaflee,trug}.}
\bea                             \label{instb1}
\mbox{monopole $(1)$: } && \left\{   \begin{array}{lcl}
n^{(1)} & \mapsto & n^{(1)}  +  2p^{(1)} \\
n^{(2)} & \mapsto &  n^{(2)}  + p^{(12)} \, ,
\end{array} \right.        \\                                   
                             \label{instb2}
\mbox{monopole $(2)$: } && \left\{   \begin{array}{lcl}
a  & \mapsto & a-N  \\
n^{(1)} & \mapsto & n^{(1)}  +  p^{(12)} \\
n^{(2)} &\mapsto  &  n^{(2)}  + 2p^{(2)} \, .
\end{array} \right.                                           
\eea   
By substituting~(\ref{instb1}) and~(\ref{instb2}) 
in~(\ref{monq}) and~(\ref{braidq}), respectively, we learn
that these local tunneling events are invisible to the long range 
monodromies with the other particles $(n^{(1)'}, n^{(2)'}, a')$
in the spectrum of this theory and that two particles 
connected by either one of these monopoles have the same 
quantum statistics phase or equivalently the same spin factor.
The conclusion is that for fixed $p^{(1)} \neq 0$ the effective 
low energy spectrum 
of the theory compactifies to  $(n^{(1)}, n^{(2)}, a)$ 
with $n^{(2)},  a \in 0,1,\ldots, N-1$ and  
$n^{(1)} \in 0,1, \ldots, 2p^{(1)}-1$. Here, it is of course understood 
that the modulo $N$ calculus for the flux quantum number $a$ 
and  the modulo $2p^{(1)}$ calculus for the charge quantum numbers
$n^{(1)}$ involve the charge jumps dispayed in~(\ref{instb1}) 
and~(\ref{instb2}) respectively. Thus all in all there are just a finite 
number $2p^{(1)} N^2$ of different stable particles in this theory.

The different 2+1 dimensional CS actions for a 
compact gauge group $G$ are known~\cite{diwi} to be classified by the 
cohomology group $H^4(BG, \Z)$. A straightforward calculation~\cite{multiCS} 
for the compact gauge group $G \simeq U(1)\times U(1)$ for example reveals 
the isomorphism $H^4(B(U(1)\times U(1)), \Z) \simeq \Z \times \Z \times \Z$. 
This is in agreement with the fact that the most general CS action
we can write down for two compact $U(1)$ gauge fields is of the 
form~(\ref{CS}). That is to say, the integral CS parameters 
$(p^{(1)}, p^{(2)}, p^{(12)})$ in~(\ref{quantmu}) label 
the different elements of $H^4(B(U(1)\times U(1)), \Z)$.
For the compact gauge group $G \simeq U(1)\times \Z_N$, in turn,
we arrive~\cite{multiCS} at the identity  
$H^4(B(U(1)\times \Z_N), \Z) \simeq \Z \times \Z_N \times \Z_N$
which indicates that two of the three integral CS parameters
in our spontaneously broken model~(\ref{action}) become cyclic with
period $N$.  An evaluation of the AB phases~(\ref{monq}) 
and~(\ref{braidq}) shows that this is indeed the case.  
To be specific,
the result of shifting $p^{(2)} \mapsto p^{(2)}+N$ in~(\ref{monq}) 
and~(\ref{braidq}) is the same as keeping $p^{(2)}$ fixed and 
replacing $(n^{(1)}, n^{(2)}, a)$ 
by $(n^{(1)}, [n^{(2)}+a], a)$ where the rectangular brackets 
denote modulo $N$ calculus in the range $0,1,\ldots, N-1$.
Something similar holds for the integral CS parameter
$p^{(12)}$. The effect of shifting $p^{(12)} \mapsto 
p^{(12)}+N$ in~(\ref{monq}) 
and~(\ref{braidq}) is the same as keeping $p^{(12)}$ fixed and 
replacing $(n^{(1)}, n^{(2)}, a)$ 
by $([n^{(1)}+a], n^{(2)}, a)$ where the rectangular brackets 
denote modulo $2p^{(1)}$ calculus in the range $0,1,\ldots,2p^{(1)}-1$.
In other words, if we just look at the long distance physics,  
then the CS parameters $p^{(2)}$ and $p^{(12)}$ are indeed cyclic
with period $N$. That is, up to a relabeling of the particles 
the theories corresponding to the integral CS parameters
$(p^{(1)}, p^{(2)}+N, p^{(12)})$ and that defined by the CS 
parameters $(p^{(1)}, p^{(2)}, p^{(12)}+N)$ describe the same spectrum 
and the same AB interactions as that with CS parameters 
$(p^{(1)}, p^{(2)}, p^{(12)})$. 
As an aside, since the number of particles in the spectrum is proportional to 
$p^{(1)}$ it is clear that this CS parameter does not exhibit any kind 
of periodicity.

\subsection{$\mu^{(1)}=0$} \label{mu0}

The partially broken CS theories~(\ref{action}) with 
$\mu^{(1)} = 0$ and $\mu^{(12)} \neq 0$ (or equivalently  
$p^{(1)}=0$ and $p^{(12)} \neq 0$) are special. 
Due to the fact that the screening flux $\phi^{(1)}$ disappears from the 
Gauss' law~(\ref{g1}), in general part of the naively expected spectrum 
becomes confined. To be concrete, after substituting 
$\mu^{(1)} = 0$ in~(\ref{g1}) along with the 
quantization~(\ref{quantmu}) of  the topological 
mass $\mu^{(12)}$ and the quantizations  $q^{(1)} = n^{(1)} e^{(1)}$
and $\phi^{(2)} = {2\pi a}/{Ne^{(2)}}$ with $n^{(1)},a \in \Z$ of 
the matter charges and the magnetic flux carried by the vortices, 
respectively, we end up  with  the condition
\bea  \label{con}
n^{(1)} = - \frac{p^{(12)}a}{N} \, .
\eea    
So in contrast to the case $\mu^{(1)} \neq 0$ the integral 
charge and flux quantum numbers $n^{(1)}$ and $a$ cease to 
be independent in this theory. It turns out to be most transparent to 
keep the flux quantum number $a$ and to consider $n^{(1)}$ as an 
`induced' screening charge.  
Since all the variables in~(\ref{con}) are integers, it is not 
guaranteed that there exists a screening charge $n^{(1)} \in \Z$ for every 
flux $a \in \Z$ such that this relation is satisfied. In fact, only if 
the flux quantum number $a$ is a multiple of
$N/\gcd(N, p^{(12)})$ (with $\gcd(N, p^{(12)})$ denoting 
the greatest common divisor of $N$ and $p^{(12)}$)
there  exists an {\em integral} screening charge $n^{(1)}$ such that 
the Gauss law~(\ref{con}) can be obeyed. 
In other words, only particles carrying magnetic flux $a$ being a 
multiple of $N/\gcd(N, p^{(12)})$ appear as free particles in 
the theory. For particles carrying flux $a$ different from multiples of 
$N/\gcd(N, p^{(12)})$, in turn, it is impossible to satisfy
the Gauss' law~(\ref{con}). From 
$(\ref{g1})$ it then follows that these particles would be 
surrounded by unscreened long range Coulomb fields $F^{(1)\, j0}$ 
corresponding to diverging Coulomb energy in this massive 
gauge theory.  The conclusion is that particles carrying fluxes $a$ different 
from multiples of $N/\gcd(N, p^{(12)})$ are confined, i.e.\ they do not 
appear as free particles.

Besides our favourite flux quantum number $a$ the particles in this theory 
may in principle be endowed with two other independent internal quantum 
numbers, namely the charge quantum number $q^{(2)} = n^{(2)} e^{(2)}$ 
with $n^{(2)} \in \Z$ and the flux quantum number $\phi^{(1)}$ which 
can take any real value as 
long as the Gauss' law~(\ref{g2}) is satisfied.
However, upon plugging the Gauss' law~(\ref{g1}) with $\mu^{(1)} = 0$
into~(\ref{mono}) and~(\ref{braid}) with $\mu^{(1)} = 0$, we see that 
the long range AB interactions are actually completely 
independent of the flux $\phi^{(1)}$ carried by the particles. 
So if we are just interested in the long distance physics of the theory 
we may nicely forget about possible fluxes $\phi^{(1)}$ attached to the 
particles and simply label these as $(n^{(2)}, a)$. With~(\ref{quantmu}) 
it is then readily checked that in terms of these two 
independent integral quantum numbers the AB phases~(\ref{mono}) 
and~(\ref{braid})  can be cast in the form
\bea
{\cal R}^2  &=& \exp \left(
 \frac{2\pi \im}{N} ( n^{(2)}a' + n^{(2)'}a + 
\frac{2p^{(2)}}{N}aa') \right),
\label{monco}  \\
{\cal R}  &=& \exp \left(
 \frac{2\pi \im}{N} ( n^{(2)}a + \frac{p^{(2)}}{N}aa) \right).  
\label{braidco}
\eea     
Since the flux quantum number $a$ is a multiple of $N/\gcd(N, p^{(12)})$, 
the integral charge label $n^{(2)}$ becomes a $\Z_{\gcd(N,p^{(12)})}$ 
quantum number under these long range interactions. Furthermore, in 
the presence of the 
minimal Dirac monopole~(\ref{instb2}) the flux quantum number $a$ is 
conserved modulo $N$. As it should, the local combined
tunnelings~(\ref{instb2}) induced by this monopole are again 
unobservable to the monodromies~(\ref{monco}) with all the particles 
in the spectrum  and two particles connected by this monopole carry the 
same quantum statistics phase or equivalently 
spin factor~(\ref{braidco}). The same obviously holds for the 
tunneling~(\ref{instb1}) with $p^{(1)}=0$ induced by the other 
minimal monopole.  Thus  the spectrum of this theory 
only consists of  a total number of $(\gcd(N, p^{(12)}))^2$ different 
stable particles which can be labeled 
as $(n^{(2)}, a)$ with $n^{(2)} \in 0,1,\ldots, \gcd(N, p^{(12)})-1$ and 
$a \in 0, N/\gcd(N, p^{(12)}), \ldots, N-N/\gcd(N, p^{(12)})$.

It is easily verified that 
the result of shifting $p^{(2)} \mapsto p^{(2)}+N$ in~(\ref{monco}) 
and~(\ref{braidco}) is the same as keeping $p^{(2)}$ fixed and 
replacing $(n^{(2)}, a)$ by $([n^{(2)}+a], a)$ where the rectangular 
brackets denote modulo $\gcd(N, p^{(12)}) $ calculus in the range 
$0,1,\ldots,\gcd(N, p^{(12)}) -1$. Hence, as in the 
previous section for $p^{(1)} \neq 0$ we infer that also for 
$p^{(1)} = 0$ the integral CS parameter $p^{(2)}$ is  
cyclic with period $N$ if we are only concerned with the long distance 
physics described by the theory. To proceed, 
the AB interactions~(\ref{monco}) and~(\ref{braidco}) are obviously  
independent of the integral CS parameter $p^{(12)}$. In fact, the only 
difference between two theories labeled by different values for 
$p^{(12)}$ is the number $\gcd(N, p^{(12)})$ of unconfined fluxes in 
the spectrum.   
Since $\gcd(N, p^{(12)}+N)=\gcd(N, p^{(12)})$ we see that the theory 
with  CS parameter $p^{(12)}$ and that with $p^{(12)}+N$
describe the same spectrum and AB interactions. 
Thus although the argument is somewhat different as that for the case  
$p^{(1)} \neq 0$ in the previous section, also for 
$p^{(1)} = 0$ the CS parameter $p^{(12)}$ 
is cyclic with period $N$ if we just consider the long 
distance physics.

For completeness' sake, for $p^{(1)}=0$ and $p^{(12)}=0$ the two compact 
$U(1)$ gauge theories obviously decouple. In the absence of monopoles the 
unbroken $U(1)$ gauge theory is in the Coulomb phase, so besides the 
screened $\Z_N$ charge/flux composites $(n^{(2)},a)$ with 
$n^{(2)},a \in 0,1, \ldots, N-1$ of the broken $U(1)$ (CS) gauge theory, 
the spectrum now also consists of quantized charges 
$q^{(1)}=n^{(1)}e^{(1)}$ with $n^{(1)} \in \Z$ carrying long range Coulomb 
fields $F^{(1)\, j0}$. In the presence of monopoles for the unbroken 
$U(1)$ gauge field, in turn, the charges $q^{(1)}=n^{(1)}e^{(1)}$ are linearly 
confined~\cite{polyakov}. Hence, the spectrum of free particles in this 
case just consists of the $\Z_N$ charge/flux composites $(n^{(2)},a)$ 
with $n^{(2)},a \in 0,1, \ldots, N-1$.

\sectiona{Concluding remarks}

In this paper we have established the spectrum and Aharonov-Bohm 
interactions for  all possible 2+1 dimensional 
Chern-Simons (CS) theories in which the compact gauge group  
$U(1) \times U(1)$ is broken down to the subgroup  
$U(1) \times \Z_N$ via the Higgs mechanism. Among other things, we found  
that the theories just featuring the mixed CS term in which the broken and 
unbroken $U(1)$ gauge fields are coupled generally exhibit an interesting 
form of confinement. Only particles carrying certain multiples of the 
fundamental vortex flux unit and certain multiples of the fundamental charge 
coupling to the unbroken $U(1)$ gauge field appear as free particles. 
Adding the standard CS term for the broken $U(1)$ gauge fields does not 
change much. It just leads to additional Aharonov-Bohm phases
among these unconfined particles. Upon adding the CS term for the 
unbroken $U(1)$ gauge fields, in contrast, the foregoing confinement 
phenomenon completely disappears from the theory. 
Along the way the tunneling properties of the Dirac monopoles that we may 
add to these compact theories were also addressed.
Moreover, it has been argued that the integral 
CS parameters labeling the mixed CS term and the standard CS term for 
the broken $U(1)$ gauge fields are cyclic with period $N$ if we are 
only considering the long distance physics described by these theories. 
No such periodicity arises for the integral 
CS parameter labeling the CS terms for the unbroken $U(1)$ gauge fields.

To conclude, with the results reported in~\cite{multiCS} the 
generalization of the above discussion to planar CS theories with more then 
two (un)broken compact $U(1)$ gauge fields is straightforward.  
An interesting question remains whether these (partially) broken compact 
abelian CS theories play a role in the setting of the fractional
quantum Hall effect.

\section*{Acknowledgements}

This work was supported by an EC grant (contract no.\ ERBCHBGCT940752).
I would like to thank Frank Wilczek for a useful conversation.

\section*{Note added}
When this work was completed a paper~\cite{corwil} appeared
which has some overlap with the discussion of section~\ref{mu0}.

\end{document}